\documentclass[amsmath,amssymb,twocolumn]{revtex4}
\usepackage{graphicx}
\makeatletter \@addtoreset{equation}{section} \makeatother

\newcommand{\disp}{\displaystyle}
\def\sech{{\rm sech}}

\def\tanh{{\rm tanh}}

\def\ri{{\rm i}}
\def\rd{{\rm d}}
\def\re{{\rm e}}
\def\PT{\mathcal{PT}}
        \newtheorem{theorem}{Theorem}
        \newtheorem{lemma}[theorem]{Lemma}

\begin{document}
\title{Nonlinear wave dynamics near phase transition in $\mathcal{PT}$-symmetric localized potentials}
\author{Sean Nixon and Jianke Yang}
\affiliation{Department of Mathematics and Statistics, University of
Vermont, Burlington, VT 05401, USA}

\begin{abstract}
Nonlinear wave propagation in parity-time ($\PT$) symmetric
localized potentials is investigated analytically near a
phase-transition point where a pair of real eigenvalues of the
potential coalesce and bifurcate into the complex plane. Necessary
conditions for phase transition to occur are derived based on a
generalization of the Krein signature. Using multi-scale
perturbation analysis, a reduced nonlinear ODE model is derived for
the amplitude of localized solutions near phase transition. Above
phase transition, this ODE model predicts a family of stable
solitons not bifurcating from linear (infinitesimal) modes under a
certain sign of nonlinearity. In addition, it predicts
periodically-oscillating nonlinear modes away from solitons. Under
the opposite sign of nonlinearity, it predicts unbounded growth of
solutions. Below phase transition, solution dynamics is predicted as
well. All analytical results are compared to direct computations of
the full system and good agreement is observed.

\end{abstract}

\maketitle

\section{Introduction}

Parity-time ($\mathcal{PT}$) symmetric systems started out from an
observation in non-Hermitian quantum mechanics, where a complex but
$\mathcal{PT}$-symmetric potential could possess all-real spectrum
\cite{Bender1998}. This concept later spread out to optics,
Bose-Einstein condensation, mechanical systems, electric circuits
and many other fields, where a judicious balancing of gain and loss
constitutes a $\mathcal{PT}$-symmetric system which can admit
all-real linear spectrum
\cite{Christodoulides2007,Guo2009,Segev2010,coupler1,Kottos2011,PT_lattice_exp,BECPT2012,Feng2013,Bender2013,Peng2014,Mercedeh2014}.
For example, in optics an even refractive index profile together
with an odd gain-loss landscape yields a $\mathcal{PT}$-symmetric
system. A common phenomenon in linear $\mathcal{PT}$-symmetric
systems is phase transition (also known as $\PT$-symmetry breaking),
where pairs of real eigenvalues collide and then bifurcate to the
complex plane when the magnitude of gain and loss is above a certain
threshold
\cite{Bender1998,Ahmed2001,Musslimani2008,coupler1,Nixon2012}. This
phase transition has been observed experimentally in a wide range of
physical systems
\cite{Guo2009,Segev2010,Kottos2011,PT_lattice_exp,Bender2013,Peng2014}.
When nonlinearity is introduced into $\mathcal{PT}$ systems, the
interplay between nonlinearity and $\mathcal{PT}$ symmetry gives
rise to additional novel properties such as the existence of
continuous families of stationary nonlinear modes, stabilization of
nonlinear modes above phase transition, and symmetry breaking of
nonlinear modes \cite{coupler1,coupler2,Musslimani2008,Konotop2011,
Nixon2012, KonotopPRL, Christodoulides_uni_2012, Panos_Dmitry2013,
SegevPT, Yang_sym_break}. These findings reveal that
$\mathcal{PT}$-symmetric systems break the boundaries between
traditional conservative and dissipative systems and open new
exciting research territories. Practical applications of
$\mathcal{PT}$ systems are starting to emerge as well, such as
recent demonstrations of $\mathcal{PT}$-symmetric micro-ring lasers
and unidirectional $\mathcal{PT}$ metamaterials
\cite{Mercedeh2014,Christodoulides_uni_2011,Christodoulides_uni_2012,Feng2013}.

An important feature of $\PT$-symmetric systems is phase transition,
where the linear spectrum changes from all-real to partially-complex
and infinitesimal waves change from stable to unstable. At phase
transition, a pair of real eigenvalues coalesce and form an
exceptional point with a non-diagonal Jordan block (i.e., with the
algebraic multiplicity higher than the geometric multiplicity).
Phase transition is a distinct linear property of $\PT$-symmetric
systems, and it is at the heart of many proposed applications such
as $\mathcal{PT}$-symmetric micro-ring lasers and unidirectional
$\mathcal{PT}$ metamaterials \cite{Mercedeh2014,Feng2013}.

When nonlinearity is present (such as if the wave amplitude is not
small), the interplay between phase transition and nonlinearity is a
fascinating subject. This interplay was previously studied for
periodic $\PT$-symmetric potentials in \cite{Nixon2012b, Nixon2013,
SegevPT}, where novel behaviors such as wave-blowup and oscillating
bound states were reported below phase transition. In addition,
stable nonlinear Bloch modes were reported above phase transition
because nonlinearity transforms the effective potential from above
to below phase transition \cite{SegevPT}. However, in periodic
potentials above phase transition, the presence of unstable
infinitely extended linear modes makes the zero background unstable,
which excludes the possibility of stable spatially-localized
coherent structures. In localized potentials, will the situation be
different?

In this article we study nonlinear wave behaviors in localized
$\PT$-symmetric potentials near phase transition. Unlike periodic
potentials, the instability of linear modes above phase transition
is limited to the area around the localized potential. In this case,
the addition of nonlinearity can balance against gain and loss
making stable spatially-localized coherent structures, such as
solitons and oscillating bound states, possible above phase
transition. Mathematically we explain this phenomenon by a
multi-scale perturbation analysis, where a reduced nonlinear ODE
model is derived for the amplitude of localized solutions near phase
transition. Above phase transition, this ODE model predicts a family
of stable solitons not bifurcating from linear (infinitesimal) modes
under a certain sign of nonlinearity. In addition, it predicts
persistent oscillating nonlinear modes away from solitons. Under the
opposite sign of nonlinearity, it predicts unbounded growth of
solutions. Similarly, solution dynamics below phase transition is
predicted as well. All these predictions are verified in the full
PDE system. In addition to these nonlinear dynamics, we also derive
a necessary condition for phase transition to occur at an
exceptional point in the linear $\PT$ system by a generalization of
the Krein signature, namely, phase transition from a collision of
two real eigenvalues is possible only when the two eigenvalues have
opposite $\PT$-Krein signatures.

\section{Preliminaries}

The mathematical model we consider in this article is the following
potential NLS equation
\begin{equation}
\ri \psi_z + \psi_{xx} +V(x; \epsilon) \psi + \sigma |\psi|^2 \psi = 0,
\label{Eq:LatticeNLS}
\end{equation}
where $V(x; \epsilon)$ is a $\mathcal{PT}$-symmetric complex
potential, i.e.,
\begin{equation}
V^*(-x;\epsilon)= V(x;\epsilon),
\end{equation}
parameterized by $\epsilon$ which controls the gain-loss strength,
$\sigma=\pm 1$ is the sign of nonlinearity, and the superscript `*'
represents complex conjugation. This model governs nonlinear light
propagation in an optical medium with gain and loss
\cite{Musslimani2008} as well as dynamics of Bose-Einstein
condensates in a double-well potential with atoms injected into one
well and removed from the other well \cite{BECPT2012}. Without loss
of generality, we assume phase transition occurs at $\epsilon=0$,
where a pair of real eigenvalues of the potential coalesce and form
an exceptional point, and we will analyze the solution dynamics in
Eq. (\ref{Eq:LatticeNLS}) near this exceptional point, i.e., when
$|\epsilon| \ll 1$.

The analysis to be developed applies to all localized
$\mathcal{PT}$-symmetric potentials near phase transition. To
illustrate these analytical results and compare them with direct
numerics of the full model (\ref{Eq:LatticeNLS}), we will use a
concrete example --- the so-called Scarff II potential
\begin{equation}
V = V_R \hspace{0.03cm} \sech^2(x) + \ri W_0 \hspace{0.03cm} \sech(x) \hspace{0.03cm} \tanh(x),
\label{Eq:ExLattice}
\end{equation}
where $V_R, W_0$ are real parameters. For this potential, phase
transition occurs at $W_0 = V_R+1/4$ \cite{Ahmed2001}, and solitons
as well as robust oscillating solutions were reported numerically
below phase transition in
\cite{Musslimani2008,Shi2011,Chen2014,Nazari2012}.

\section{$\mathcal{PT}$-Krein Signature and a Necessary Condition for Phase Transition}

We begin by studying the general linear eigenvalue problem
\begin{equation} \label{Lu}
L(x;\epsilon) u = -\mu u,
\end{equation}
where $L$ is a $\PT$-symmetric linear operator parameterized by
$\epsilon$, i.e.,
\begin{equation}
L^*(-x;\epsilon)= L(x;\epsilon),
\end{equation}
and $\mu$ is an eigenvalue. We wish to consider the phase-transition
process by which the spectrum of $L$ changes from all-real to
partially-complex. This phase transition occurs when a pair of real
eigenvalues collide and then bifurcate into the complex plane. It is
important to recognize that not any two real eigenvalues can turn
complex upon collision. This is analogous to the linear stability of
equilibria in Hamiltonian systems, where not just any two purely
imaginary eigenvalues upon collision can bifurcate off the imaginary
axis and create linear instability
\cite{MacKay,KapitulaBook,Pelinovsky_book}. Then the question we
address is: under what conditions can a pair of real eigenvalues of
$L$ induce phase transition upon collision?

For the potential NLS equation (\ref{Eq:LatticeNLS}), when one looks
for linear eigenmodes $\psi = u(x) \re^{-\ri \mu z}$, the eigenvalue
problem (\ref{Lu}) will be obtained with
\begin{equation}
L = \partial_{xx} + V(x;\epsilon),
\label{e:Eig}
\end{equation}
which is $\PT$-symmetric. However, in this section we will consider
the eigenvalue problem (\ref{Lu}) for general $\PT$-symmetric
operators, not just (\ref{e:Eig}).

First we make some elementary observations. Since the operator $L$
is $\mathcal{PT}$-symmetric, any complex eigenvalues must come in
conjugate pairs, i.e., if $[\mu, u(x)]$ is an eigenmode, then so is
$[\mu^*, u^*(-x)]$. If $\mu$ is a simple real eigenvalue, then its
eigenfunction $u(x)$ can be made $\PT$-symmetric by scaling.

We start the analysis by introducing a sesquilinear $\PT$-product
\begin{equation}
\langle f, g\rangle_{\PT} \equiv \int_{-\infty}^{\infty} f^*(-x) g(x) \rd x,
\label{e:PTproduct}
\end{equation}
which naturally satisfies the symmetry condition
\begin{equation} \label{e:gf}
\langle g, f \rangle_{\PT}  = \langle f, g\rangle_{\PT}^*.
\end{equation}
Thus for any complex function $f(x)$, $\langle f, f \rangle_{\PT}$
is real and invariant under a gauge transformation $f(x)\to
f(x)e^{i\theta}$, where $\theta$ is a real constant. In addition,
under this $\PT$-product the $\PT$-symmetric operator $L$ is
``self-adjoint", i.e.,
\begin{equation}  \label{e:Lfg}
\langle Lf, g\rangle_{\PT}=\langle f, Lg\rangle_{\PT}.
\end{equation}

For an eigenmode $[\mu, u(x)]$ of \eqref{e:Eig} with a simple real
eigenvalue $\mu$ we define its $\PT$-Krein signature as
\begin{equation}  \label{def:S}
{\rm S}(\mu) = {\rm sgn} \left[\langle u, u\rangle_{\PT}  \right].
\end{equation}
Here $\langle u, u\rangle_{\PT} \ne 0$, because $u(x)$ can be made
$\PT$-symmetric, i.e., $u^*(-x)=u(x)$, so $\langle u,
u\rangle_{\PT}=\int_{-\infty}^\infty u^2dx$, which is nonzero since
it is the Fredholm condition for the generalized-eigenfunction
equation $(L+\mu)u_g=u$ not to admit a solution in view that
$u^*(x)$ is in the kernel of the adjoint operator $L^*+\mu$. Thus
the $\PT$-Krein signature of a simple real eigenvalue is always
positive or negative. In addition, this signature cannot change
under continuous variation of the parameter $\epsilon$ unless pairs
of such eigenvalues collide.

The main result of this section is that when two such eigenvalues
collide, a necessary condition for complex-eigenvalue bifurcation is
the two real eigenvalues have opposite $\PT$-Krein signatures. This
result extends an analogous one in Hamiltonian systems to
$\PT$-symmetric systems \cite{MacKay,KapitulaBook}.

We first present three lemmas.
\begin{lemma}\label{Lemma: Orthogonality}
Let $u_1(x)$ and $u_2(x)$ be two eigenfunctions of the
$\PT$-symmetric operator $L$ with real eigenvalues $\mu_1$ and
$\mu_2$ respectively. If $\mu_1\ne \mu_2$ then $\langle u_1,
u_2\rangle_{\PT} =0$.
\end{lemma}

\noindent\textbf{Proof}\ Using the ``self-adjoint" property
(\ref{e:Lfg}) we have
\[
\langle u_1, Lu_2\rangle_{\PT}  = \langle Lu_1, u_2\rangle_{\PT}.
\]
Then using the fact that $u_1$ and $u_2$ are eigenfunctions, we can
calculate the left and right sides of the above equation as
\[\langle u_1, Lu_2\rangle_{\PT} = - \mu_2 \langle u_1, u_2\rangle_{\PT},\]
and
\[
\langle Lu_1, u_2\rangle_{\PT}=- \mu_1 \langle u_1, u_2\rangle_{\PT}.
\]
Thus if $\mu_1\ne \mu_2$ then $\langle u_1, u_2\rangle_{\PT} =0$.
$\Box$

\begin{lemma}\label{Lemma: Orthogonality2}
Let $u(x)$ be an eigenfunction of the $\PT$-symmetric operator $L$
with a complex eigenvalue $\mu$. Then $\langle u, u\rangle_{\PT}
=0$.
\end{lemma}

\noindent \textbf{Proof}\  From Eq. (\ref{e:Lfg}) we have $\langle
u, Lu\rangle_{\PT} = \langle Lu, u\rangle_{\PT}$. Calculating the
two sides of this equation we get
\[
(\mu-\mu^*)\langle u, u\rangle_{\PT}=0.
\]
Thus if $\mu$ is complex, then $\langle u, u\rangle_{\PT} =0$.
$\Box$

\begin{lemma}
Let $\{ e_j \}$ be a basis for an $N$-dimensional functional
subspace and $f = \disp\sum_{j=1}^N c_j e_j$. Then
\begin{equation*}
\langle f, f\rangle_{\PT} = c^HMc,
\end{equation*}
where $M$ is a $N\times N$ Hermitian matrix with elements given by
\begin{equation}
M_{ij} = \langle e_i, e_j\rangle_{\PT},
\end{equation}
and the superscript `H' represents the Hermitian of a vector.
\end{lemma}
\textbf{Proof}\ Substituting the $f$ expression into $\langle f,
f\rangle_{\PT}$ and utilizing the linearity of the $\PT$-product,
this lemma can be readily proved. The Hermiticity of $M$ comes
directly from relations (\ref{e:gf}) and (\ref{e:Lfg}). $\Box$

Now we present the main result of this section.
\begin{theorem} \label{Theorem1}
Let $L$ be a $\PT$-symmetric operator parameterized by $\epsilon$.
If a pair of simple real eigenvalues of $L$ collide and bifurcate
into the complex plane at $\epsilon = 0$, then before the
bifurcation the two real eigenvalues must have opposite $\PT$-Krein
signatures.
\end{theorem}

\textbf{Proof}\ Let $u_1(x)$ and $u_2(x)$ be two eigenfunctions of
$L$ with eigenvalues $\mu_1$ and $\mu_2$ respectively; when
$\epsilon<0$, $\mu_1$ and $\mu_2$ are simple real with $\mu_1\neq
\mu_2$, and when $\epsilon>0$, $\mu_1$ and $\mu_2$ are complex with
$\mu_1 = \mu_2^*$. We analyze the quadratic form $\langle f,
f\rangle_{\PT}$ restricted to the subspace $S = {\rm span}(u_1,u_2)$
by looking at the dual matrix $M$.

When $\epsilon>0$ and $\mu_1 = \mu_2^*$, we see easily by Lemma
\ref{Lemma: Orthogonality2} that
\begin{equation}
M_{\epsilon>0} = \left[ \begin{array}{cc}  0  & b \\ b^* & 0  \end{array} \right],
\end{equation}
where $b =  \langle u_1, u_2\rangle_{\PT}$. In this case,
$u_1(x)=u_2^*(-x)$, thus
\[b=\langle u_2^*(-x), u_2(x)\rangle_{\PT}=\int_{-\infty}^\infty u_2^2(x)dx, \]
which is nonzero because for a simple eigenvalue $\mu_2$,
$\int_{-\infty}^\infty u_2^2dx\ne 0$ is the Fredholm condition for
non-existence of a generalized eigenfunction [see earlier text below
Eq. (\ref{def:S})]. This means $M$ has a pair of real eigenvalues of
opposite sign and is thus indefinite.

Likewise, for $\epsilon<0$, since $\mu_1$ and $\mu_2$ are strictly
real, then in view of Lemma \ref{Lemma: Orthogonality},
\begin{equation} \label{Mepsin}
M_{\epsilon<0} = \left[ \begin{array}{cc}  a_1 & 0 \\ 0 & a_2 \end{array} \right],
\end{equation}
where $a_1 = \langle u_1, u_1\rangle_{\PT} $ and $a_2 = \langle u_2,
u_2\rangle_{\PT}$ have the signs of the $\PT$-Krein signatures for
$\mu_1$ and $\mu_2$ respectively.

At $\epsilon = 0$ the two eigenvalues collide, and $\mu_1 = \mu_2 =
\mu_0$.  There are two cases here. The first case is where $L$ has
two eigenfunctions $u_1$ and $u_2$ at $\mu_0$, i.e., $L$ has a
diagonal Jordan block. In this case, $\langle u_k,
u_k\rangle_{\PT}\ne 0$ ($k=1, 2$), and we can always choose the
basis $\{u_1, u_2\}$ so that $\langle u_1, u_2\rangle_{\PT}=0$, thus
the operator $M_{\epsilon=0}$ has the same structure as
(\ref{Mepsin}). The second case is where $L$ has a single
eigenfunction $u_0$ at $\mu_0$, i.e., $L$ has a non-diagonal Jordan
block. In this case, the subspace $S$ reduces to $S = {\rm
span}(u_0,u_g)$, where $u_g$ is the generalized eigenfunction
satisfying
\begin{equation*}
(\mu_0+L) u_g = u_0.
\end{equation*}
Taking the $\PT$-product of this equation with $u_0$ and recalling
(\ref{e:Lfg}), i.e., applying the Fredholm condition, we get
$\langle u_0, u_0\rangle_{\PT} =0$, thus
\begin{equation}
M_{\epsilon=0}  = \left[ \begin{array}{cc}  0 & c_1 \\ c_1^* & c_2 \end{array} \right],
\end{equation}
where $ c_1 = \langle u_0, u_g\rangle_{\PT}$ and $c_2 = \langle u_g,
u_g\rangle_{\PT}$. Note that $c_1\ne 0$ since it is the Fredholm
condition for eigenvalue $\mu_0$ not to have a second generalized
eigenfunction. Then det$(M_{\epsilon=0})<0$, hence $M_{\epsilon=0}$
has a pair of real eigenvalues of opposite sign and is indefinite.

Since $M$ is indefinite for $\epsilon > 0$ and this indefiniteness
is continuous across the bifurcation point $\epsilon=0$, we see that
$M$ for $\epsilon < 0$ must also be indefinite, which directly
implies that real eigenvalues $\mu_1$ and $\mu_2$ before bifurcation
must have opposite $\PT$-Krein signatures. $\Box$

\begin{figure}[!htbp]
    \centering
    \includegraphics[width=0.48\textwidth]{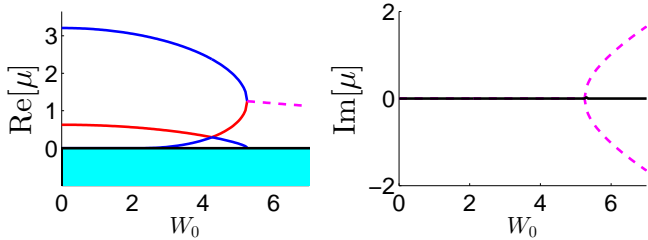}
    \caption{Eigenvalues $\mu$ for varying gain-loss strength $W_0$ in the Scarff-II potential (\ref{Eq:ExLattice}) with $V_R=5$
    (the continuous spectrum is displayed in light blue). Blue lines indicate eigenvalues with positive $\PT$-Krein signatures
    while red lines indicate eigenvalues with negative $\PT$-Krein signatures.
    Complex eigenvalues are indicated with dotted purple lines. }
    \label{Fig:PTKrien}
\end{figure}

Now we use an example to illustrate this theorem. In the Scarff-II
potential (\ref{Eq:ExLattice}), we fix $V_R=5$ and vary the
gain-loss coefficient $W_0$. The linear spectra for various $W_0$
values are displayed in Fig.~\ref{Fig:PTKrien}. It is seen that
phase transition occurs at $W_0=5.25$, where a pair of simple real
eigenvalues coalesce and form an exceptional point, which then turns
complex when $W_0>5.25$. We have calculated the $\PT$-Krein
signatures of those real eigenvalues (indicated by colors in the
figure) and found them indeed opposite, in agreement with
Theorem~\ref{Theorem1}.

Interestingly, Fig.~\ref{Fig:PTKrien} also shows another collision
of simple real eigenvalues of opposite $\PT$-Krein signatures at
$W_0=4.25$ (which creates another exceptional point). However, after
collision these real eigenvalues re-emerge and no complex
eigenvalues bifurcate out. This shows that collision of real
eigenvalues of opposite $\PT$-Krein signatures is a necessary but
not sufficient condition for complex-eigenvalue bifurcation. A
sufficient condition for complex-eigenvalue bifurcation can be found
in later text [i.e., $\alpha\ne 0$, see the paragraph below Eq.
(\ref{e:I2})]. This condition is not met at the other exceptional
point, thus we do not see phase transition there.

\section{Reduced Model Near Phase Transition}

In this section, we consider the potential nonlinear Schr\"{o}dinger
equation \eqref{Eq:LatticeNLS} and analyze its solution dynamics
near phase transition.

Let us suppose the $\PT$-symmetric potential in
Eq.~\eqref{Eq:LatticeNLS} takes the form
\begin{equation}  \label{e:V}
V(x; \epsilon) = V_0(x) + \epsilon^2 V_2(x),
\end{equation}
where $0<\epsilon\ll 1$. Here $V_0(x)$ is the unperturbed potential,
$V_2(x)$ is the form of potential perturbation, and $\epsilon^2$ is
the strength of this perturbation. We assume that the unperturbed
potential $V_0(x)$ is at phase transition and possesses an
exceptional point at $\mu=\mu_0$, i.e., the linear operator
$\partial_{xx}  + V_0(x)$ has a single eigenfunction $u_e(x)$ and a
generalized eigenfunction $u_g(x)$ at $\mu_0$. Defining
\begin{equation}
L_0 \equiv  \partial_{xx}  + V_0(x) +\mu_0,
\end{equation}
then we have
\begin{equation} \label{L0ueug}
L_0u_e=0, \quad L_0u_g=u_e.
\end{equation}
Since $L_0$ is $\PT$-symmetric, both $u_e$ and $u_g$ can be chosen
to be $\PT$-symmetric as well, i.e.,
\begin{equation}
u_e^*(-x)=u_e(x), \quad   u_g^*(-x)=u_g(x).
\end{equation}
By taking the complex conjugate of the $L_0u_e=0$ equation, we see
that $u_e^*$ is in the kernel of the adjoint operator $L_0^*$. Thus
the solvability condition for the $L_0u_g=u_e$ equation is that
\begin{equation} \label{e:ue2}
\int_{-\infty}^{\infty} u_e^2 \rd x = 0.
\end{equation}

In principle, an exceptional point can have algebraic multiplicities
higher than two, meaning that it can have additional generalized
eigenfunctions beside $u_g$. But in a generic case, an exceptional
point is formed by the collision of two simple real eigenvalues, in
which case its algebraic multiplicity is only two. For simplicity,
we only consider such generic exceptional points in this article.
Since their algebraic multiplicities are two, they do not admit
other generalized eigenfunctions, i.e., the equation
\begin{equation*}
L_0u_{g2}=u_g
\end{equation*}
admits no localized solutions for $u_{g2}$. Since $u_e^*$ is in the
kernel of the adjoint operator $L_0^*$, the Fredholm condition on
the above equation is that its right side be not orthogonal to
$u_e^*$, i.e.,
\begin{equation}
D\equiv  \int_{-\infty}^{\infty} u_e u_g \rd x \ne 0.
\end{equation}
In addition, since $u_e$ and $u_g$ are $\PT$-symmetric, so is
$u_eu_g$, hence $D$ is real.

If the potential $V_0(x)$ is perturbed to be (\ref{e:V}), we study
nonlinear dynamics in this perturbed potential by multiscale
perturbation methods below. First we expand the solution to
Eq.~\eqref{Eq:LatticeNLS} into a perturbation series,
\begin{equation} \label{e:psiexpansion}
\psi(x, z) = \left( \epsilon u_1(x, Z) + \epsilon^2 u_2 +
\epsilon^3 u_3 +\ldots\right)\re^{-\ri \mu_0 z},
\end{equation}
where $Z = \epsilon z$. Then up to order $\epsilon^3$ we have a
system of equations
\begin{align*}
L_0 u_1 &= 0, \\
L_0 u_2 & = -\ri u_{1Z},\\
L_0 u_3 & = -\ri u_{2Z} - V_2 u_1 - \sigma |u_1|^2 u_1.
\end{align*}
Since $u_e^*$ is in the kernel of the adjoint operator $L_0^*$, the
solvability conditions for these equations are that their right
sides be orthogonal to the adjoint homogeneous solution $u_e^*$.

At orders $\epsilon$ and $\epsilon^2$ we find from (\ref{L0ueug})
that
\begin{subequations} \label{e:u1u2}
\begin{align}
u_1 & = A(Z) u_e(x), \\
u_2 & = -\ri A_{Z} u_g.
\end{align}
\end{subequations}
At order $\epsilon^3$ we have
\begin{equation*}
L_0 u_3 = - A_{ZZ} u_g - A V_2 u_e - \sigma |A|^2 A |u_e|^2 u_e.
\end{equation*}
The solvability condition of this equation is
\begin{equation}
A_{ZZ} - \alpha A + \sigma_1 |A|^2 A = 0,
\label{Eq:AFull}
\end{equation}
where
\begin{equation} \label{e:alphasigma1}
\alpha =  - \frac{1}{D} \int_{-\infty}^{\infty} V_2 u_e^2 \rd x, \quad
\sigma_1 =  \frac{\sigma}{D} \int_{-\infty}^{\infty} |u_e|^2 u_e^2 \rd x.
\end{equation}
Equation (\ref{Eq:AFull}) for the wave envelope $A(Z)$ is our
reduced model for nonlinear wave dynamics near an exceptional
(phase-transition) point. Since $V_2$ and $u_e$ are $\PT$-symmetric
and $D$ real, $\alpha$ and $\sigma_1$ are real.

The reduced model (\ref{Eq:AFull}) is a fourth-order dynamical
system since $A$ is complex. However it has two conserved
quantities,
\begin{equation}
I_1=|A_Z|^2-\alpha |A|^2+\frac{\sigma_1}{2}|A|^4,
\end{equation}
and
\begin{equation} \label{e:I2}
I_2=A^*A_Z-AA_Z^*,
\end{equation}
where $dI_k/dZ=0$ ($k=1, 2$). Due to these two conserved quantities,
solution dynamics in Eq. (\ref{Eq:AFull}) is confined to a
two-dimensional surface, thus this dynamics cannot be chaotic. When
$Z\to \infty$, the solution $A$ can only approach a fixed point, or
a periodic orbit, or infinity (if $\sigma_1>0$, infinity is further
forbidden due to conservation of $I_1$).

The parameter $\alpha$ plays an important role in
Eq.~(\ref{Eq:AFull}). Let us consider the small-amplitude limit
($|A|\ll 1$), in which case Eq.~(\ref{Eq:AFull}) reduces to
$A_{ZZ}-\alpha Z=0$. If $\alpha <0$, these infinitesimal (linear)
modes are bounded, meaning that the system is below phase
transition. But, if $\alpha>0$, these linear modes exponentially
grow, indicating that the system is above phase transition. Recall
that $\alpha$ is dependent on the potential perturbation $V_2$. Thus
whether the perturbed potential is above or below phase transition
depends on the sign of $\alpha$. In addition, the value of $\alpha$
also determines whether or not the underlying exceptional point
$\mu=\mu_0$ is a phase-transition point: if $\alpha\ne 0$, then
$\mu=\mu_0$ is a phase-transition point; if $\alpha=0$, then the
answer is not certain, and further analysis is needed in order to
determine whether $\mu=\mu_0$ is a phase-transition point or not.

In the next two sections, we will describe the predictions of the
reduced model (\ref{Eq:AFull}) and compare them with the full system
(\ref{Eq:LatticeNLS}). In all our numerical comparisons, we will use
the Scarff-II potential (\ref{Eq:ExLattice}) with $V_R=2$. At this
$V_R$ value, an exceptional point occurs when
\begin{equation}
W_0=2.25, \quad \mu_0\approx -0.3144,
\end{equation}
and this exceptional point is a phase-transition point. In the
format (\ref{e:V}) of the perturbed potential, this Scarff-II
potential has
\begin{subequations} \label{e:ScarffII}
\begin{align}
V_0(x) &= 2 \sech^2(x) + \ri \hspace{0.03cm} 2.25 \hspace{0.03cm} \sech(x) \hspace{0.03cm} \tanh(x), \\
V_2(x) & = \ri \hspace{0.03cm} c \hspace{0.04cm} \sech(x) \hspace{0.03cm} \tanh(x).
\end{align}
\end{subequations}
This potential is above phase transition when $c = 1$ and below
phase transition when $c=-1$. At this phase-transition point, the
coefficients in the reduced model (\ref{Eq:AFull}) are found to be
\begin{equation}  \label{e:alphasigma}
\alpha \approx 0.3144c, \quad \sigma_1 \approx 0.2700\sigma.
\end{equation}
For these coefficients the eigenfunction $u_e(x)$ has been
normalized to have unit amplitude. In all our comparisons, we take
$\epsilon=0.2$. This $\epsilon$ is not very small, but predictions
of the reduced model (\ref{Eq:AFull}) still match those in the full
system (\ref{Eq:LatticeNLS}) as we will see below.

\section{Solution Behaviors Above Phase Transition}

Our main interest is to investigate nonlinear wave dynamics above
phase transition ($\alpha>0$). Previous studies on nonlinear
$\PT$-symmetric systems overwhelmingly focused on solution behaviors
below phase transition, because it was argued that coherent
structures such as solitons would be unstable above phase transition
(at least in $\PT$-symmetric periodic potentials). We will show in
this section that in $\PT$-symmetric localized potentials, stable
solitons and robust oscillating nonlinear modes do exist above phase
transition.

\subsection{Soliton families and their stability}

First we consider soliton solutions, which correspond to
constant-amplitude solutions in the reduced model \eqref{Eq:AFull}.
Specifically, constant-amplitude solutions of the form
\begin{equation} \label{e:AZ}
A(Z) = A_0 \re^{- \ri \mu_1 Z}
\end{equation}
in Eq.~\eqref{Eq:AFull} correspond to soliton solutions of the form
\begin{equation} \label{e:psisoliton}
\psi = u(x) \re^{-\ri (\mu_0 + \epsilon \mu_1) z}
\end{equation}
in Eq.~(\ref{Eq:LatticeNLS}), where $u(x)\approx \epsilon A_0u_e(x)$
to leading order. In this $A$-formula, $A_0$ will be made real
positive from phase invariance. Substituting (\ref{e:AZ}) into
\eqref{Eq:AFull}, we find $\mu_1$ as
\begin{equation} \label{e:mu1}
\mu_1 =\pm \sqrt{\sigma_1 A_0^2 - \alpha},
\end{equation}
where the quantity under the square root must be non-negative. This
equation relates the propagation constant $\mu_1$ to the soliton
amplitude parameter $A_0$.

Since $\alpha>0$ above phase transition, solutions (\ref{e:mu1})
exist only when $\sigma_1>0$. For the Scarff-II potential
(\ref{e:ScarffII}), this means that above phase transition, solitons
can only exist under self-focusing nonlinearity ($\sigma>0$). The
physical reason for the existence of these solitons comes from the
nonlinear feedback. It is commonly known that a $\PT$-symmetric
complex potential is above phase transition when the imaginary part
of the potential (relative to the real part) is above a certain
threshold. In the current case, even though the linear potential
$V(x;\epsilon)$ is above phase transition, the nonlinearity-induced
positive refractive index $\sigma|\psi|^2$, when added to this
linear potential, enhances its real part, which makes its imaginary
part relatively weaker. As a consequence, the nonlinearity
transforms the effective potential from above phase transition to
below phase transition \cite{SegevPT}.

Notice also from Eq. (\ref{e:mu1}) that these solitons exist only
above a certain amplitude (or power) threshold, which is
\begin{equation}  \label{e:A02}
A_0^2\ge \alpha/\sigma_1.
\end{equation}
This means that the nonlinearity-induced positive refractive index
must be strong enough in order to transform the effective potential
from above to below phase transition. Consequently, these solitons
do not bifurcate from linear modes of the potential. In addition,
the two branches of these solutions [corresponding to the plus and
minus signs in (\ref{e:mu1})] are connected at this amplitude
threshold and thus belong to a single soliton family.

\begin{figure}[!htbp]
    \centering
    \includegraphics[width=0.48\textwidth]{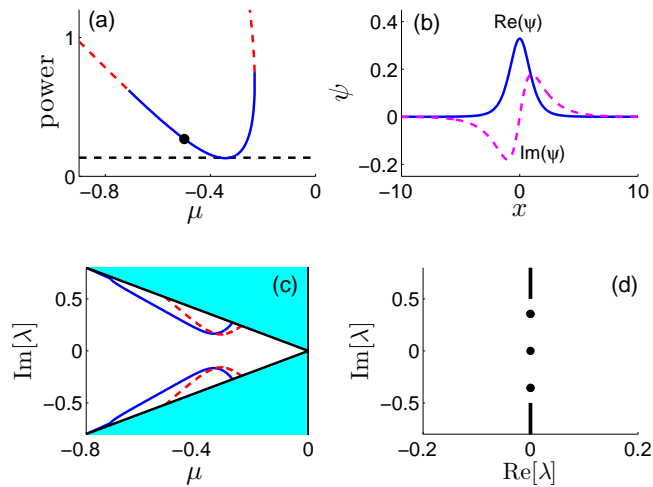}
    \caption{(a) Power curve for the family of solitons above phase transition in the Scarff-II potential (\ref{e:ScarffII}) with $\sigma=1$, $c=1$ and $\epsilon=0.2$; solid blue
    indicates stable solitons and dashed red unstable ones; the horizontal dashed line is the analytical prediction for the power minimum;
    (b) profile of an example soliton at the black-dot point of the power curve (where $\mu=-0.5$); (c) discrete eigenvalues for stable solitons versus the propagation constant $\mu$;
    solid blue are numerical values and dashed red analytical predictions; the shaded region is the continuous spectrum; (d) the numerically
    obtained linear-stability spectrum for the soliton in panel~(b). }
    \label{Fig2}
\end{figure}

For the Scarff-II potential (\ref{e:ScarffII}) with $\sigma=1$,
$c=1$ and $\epsilon=0.2$, we have numerically obtained these
predicted solitons above phase transition, whose power curve is
plotted in Fig.~\ref{Fig2}(a). Here the soliton's power is defined
as $P(\mu)=\int_{-\infty}^\infty |\psi|^2 dx$. It is seen that this
numerical power curve indeed has a minimum threshold. The analytical
prediction for this power threshold, obtained from Eq. (\ref{e:A02})
and the leading-order perturbation solution (\ref{e:psiexpansion})
as
\[
P_{min}=\frac{\alpha \epsilon^2}{\sigma_1}
\int_{-\infty}^\infty |u_e|^2dx,
\]
is also depicted in Fig.~\ref{Fig2}(a) (as a horizontal dashed
line). It is seen that this analytical power threshold matches the
numerical value very well. At the black-dot point of the numerical
power curve (where $\mu=-0.5$), the profile of the corresponding
soliton solution is illustrated in Fig.~\ref{Fig2}(b). This soliton
is $\PT$-symmetric, as are all other solitons in this family.

Stability of these solitons can be analyzed by examining the
stability of constant-amplitude solutions (\ref{e:AZ}) in the
reduced ODE model (\ref{Eq:AFull}). Let us perturb this
constant-amplitude solution by normal modes as
\[
A(Z) = \left( A_0 + \widetilde{A} \hspace{0.05cm} \re^{ \lambda_A Z} +\widetilde{B}^* \re^{
\lambda_A^* Z}\right) \re^{-\ri \mu_1 Z},
\]
where $\widetilde{A}, \widetilde{B}\ll 1$, and $\lambda_A$ is the
eigenvalue from the envelope equation. Plugging this into
\eqref{Eq:AFull} and linearizing, we obtain
\begin{equation*}
L_A \left(\begin{array}{c} \widetilde{A}\\ \widetilde{B} \end{array} \right) = 0,
\end{equation*}
where
\begin{equation*}
L_A = \left( \begin{array}{c c} \lambda_A^2 - 2\ri \lambda_A\mu_1 + \mu_1^2+\alpha   &  \mu_1^2+\alpha \\
\mu_1^2+\alpha &\lambda_A^2 + 2\ri \lambda_A\mu_1 + \mu_1^2+\alpha\end{array} \right).
\end{equation*}
Requiring the determinant of this matrix $L_A$ to vanish, non-zero
eigenvalues $\lambda_A$ are then derived as
\[
\lambda_A = \pm i\sqrt{2(3\mu_1^2+\alpha)}.
\]
Since $\alpha>0$ above phase transition, this formula predicts a
pair of purely imaginary discrete eigenvalues, indicating that the
constant-amplitude solution (\ref{e:AZ}) is stable in the ODE model
(\ref{Eq:AFull}). This implies that the soliton solution
(\ref{e:psisoliton}) is also stable in the original model
(\ref{Eq:LatticeNLS}). Taking into account the scaling $Z=\epsilon
z$, an approximation for non-zero discrete eigenvalues of this
soliton is
\begin{equation}  \label{e:lambda}
\lambda \approx \epsilon \lambda_A = \pm i\epsilon \sqrt{2(3\mu_1^2+\alpha)}.
\end{equation}

Numerically we have confirmed the stability of these solitons near
phase transition. This is achieved by computing the linear-stability
spectrum of these solitons by the Fourier-collocation method
\cite{Yang_SIAM}. For example, for the soliton shown in
Fig.~\ref{Fig2}(b), its linear-stability spectrum is displayed in
Fig.~\ref{Fig2}(d). All eigenvalues in this spectrum are purely
imaginary, indicating that the soliton is linearly stable. In
addition, the pair of discrete imaginary eigenvalues in this
spectrum correspond to those predicted analytically by formula
(\ref{e:lambda}). Similar computations are performed for other
solitons, and their stability is indicated by solid blue lines on
the power diagram of Fig.~\ref{Fig2}(a). Quantitative comparison
between numerical discrete imaginary eigenvalues and their
analytical prediction (\ref{e:lambda}) is made in
Fig.~\ref{Fig2}(c), and reasonable agreement can be seen (even
though $\epsilon=0.2$ is not small here).

At high powers, we find that these solitons above phase transition
become linearly unstable, and this instability is shown on the power
curve of Fig.~\ref{Fig2}(a) by dashed red lines. The instability on
the left side of the power curve is induced by complex-eigenvalue
bifurcations from edges of the continuous spectrum, while
instability on the right side of the power curve is caused by
complex-eigenvalue bifurcations from interiors of the continuous
spectrum. These high-power solitons correspond to amplitude values
on the order $A\sim1/\epsilon$ and lie outside the validity of our
perturbation theory, thus their instability does not contradict our
stability result for low-power solitons.

\subsection{Oscillating solutions}

The behavior of solutions away from the soliton equilibriums can be
largely captured by focusing on the case where $A$ is purely real in
the reduced system \eqref{Eq:AFull}. In this case, the model
equation becomes a simple second order ODE which we analyze using
phase portraits. Above phase transition, $\alpha>0$, this breaks
into two cases depending on the sign of the nonlinearity.

\subsubsection{Positive $\sigma_1$}

\begin{figure}[!htbp]
    \centering
    \includegraphics[width=0.48\textwidth]{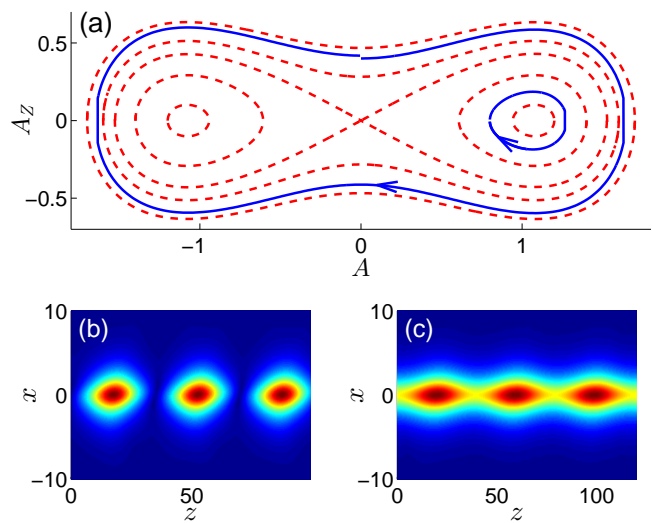}
    \caption{(a) Phase portrait of the reduced model \eqref{Eq:AFull} with $\sigma_1>0$ above phase transition. (b,c) Full PDE simulations
    in the Scarff-II potential (\ref{e:ScarffII}) with $\sigma=1$, $c=1$ and
    $\epsilon=0.2$ under different initial conditions. In (a), solid blue lines are full
    PDE solutions in (b,c) projected onto the phase plane [with the outer curve for (b) and inner curve for (c)]. }
    \label{Fig3}
\end{figure}

In this case, the phase portrait is shown in Fig.~\ref{Fig3}(a),
where $\alpha$ and $\sigma_1$ values are taken from Eq.
(\ref{e:alphasigma}) with $c=1$ and $\sigma=1$ (focusing
nonlinearity). This phase portrait contains three fixed points. One
of them is the origin, which is unstable, signifying that the system
is above phase transition. The other two fixed points are at $A=\pm
\sqrt{\alpha/\sigma_1}$, which are stable, and they correspond to
the soliton of minimum power (with $\mu_1=0$) in Eqs.
(\ref{e:mu1})-(\ref{e:A02}). Away from these fixed points, the phase
portrait features two types of periodic orbits which are separated
by a figure-eight trajectory joined at the origin. Inner periodic
orbits surround the non-zero fixed points, while outer periodic
orbits undergo wider amplitude swings.

These periodic orbits in the phase plane imply the existence of
robust oscillating solutions away from solitons in the full PDE
\eqref{Eq:LatticeNLS}, and such oscillating solutions are confirmed
in our direct evolution simulations of that system. To illustrate,
two examples of such PDE solutions are displayed in
Fig.~\ref{Fig3}(b,c). Oscillations in panel (b) are stronger, and
they correspond to outer periodic orbits in the phase portrait (a).
Oscillations in panel (c) are weaker, and they correspond to inner
periodic orbits in the phase portrait. This solution correspondence
can be made more explicit by projecting the PDE solution onto the
phase plane. To do so, we recall the perturbation solution
(\ref{e:psiexpansion})-(\ref{e:u1u2}), which to order $\epsilon^2$
gives
\begin{equation}  \label{e:psiproj}
\psi(x,z) = \left[\epsilon A(Z) u_e(x) - \ri \epsilon^2 A'(Z) u_g(x)\right]\re^{-\ri \mu_0 z}.
\end{equation}
Taking the inner product of this equation with $u_g(x)$ and
retaining only the leading-order term, we get
\begin{equation} \label{e:AZG}
A(Z)= \frac{1}{\epsilon D} \int_{-\infty}^{\infty} \psi(x, Z/\epsilon) \hspace{0.04cm} u_g(x) \hspace{0.06cm} \rd \hspace{-0.03cm} x \,\, \re^{\ri \mu_0 Z/\epsilon}.
\end{equation}
Taking the inner product of (\ref{e:psiproj}) with $u_e(x)$ and
recalling the relation (\ref{e:ue2}), we get
\begin{equation} \label{e:AprimeZ}
A'(Z)=\frac{\ri}{\epsilon^2 D} \int_{-\infty}^{\infty}  \psi(x, Z/\epsilon) \hspace{0.04cm} u_e(x) \hspace{0.06cm} \rd \hspace{-0.03cm} x \,\,
\re^{\ri \mu_0 Z/\epsilon}.
\end{equation}
In this way, the full PDE solution can be embedded in the phase
portrait of the ODE model for comparison. As a technical matter, the
projected quantities $(A, A')$ from the PDE solution by
(\ref{e:AZG})-(\ref{e:AprimeZ}) are complex in general. But we have
found that if the initial condition of the PDE solution is chosen
according to Eq. (\ref{e:psiproj}), then the imaginary parts of the
projected $(A, A')$ remain very small for very long distances. Thus
we neglect those small imaginary parts and plot only the real parts
of the projected $(A, A')$ in the phase plane.

For the two PDE solutions in Fig.~\ref{Fig3}(b,c), their phase-plane
projections are displayed as solid blue lines in panel (a). It is
seen that these PDE projections closely mimic the periodic orbits of
the ODE model.

It is noted that these predictions of periodically-oscillating
solutions in the PDE system are valid on the distance scale of
$z\sim 1/\epsilon$. Beyond this distance scale, the PDE dynamics
generally starts to deviate from the ODE predictions. Our numerics
shows that over very long distances, these oscillations in the PDE
solution gradually intensify and eventually break up, which is
caused by resonance of nonlinearity-induced higher harmonics of
these oscillations with the continuous spectrum in our opinion.

\subsubsection{Negative $\sigma_1$}

\begin{figure}[!htbp]
    \centering
    \includegraphics[width=0.48\textwidth]{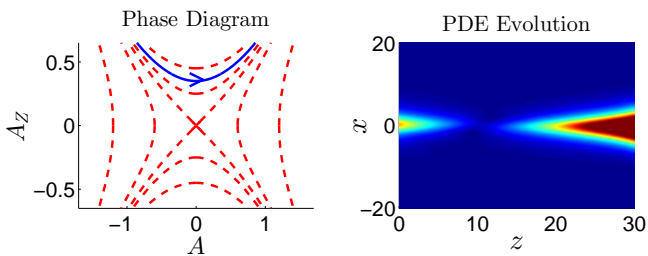}
    \caption{ (Left) Phase portrait of the reduced model \eqref{Eq:AFull} with $\sigma_1<0$ above phase transition. (Right) Full PDE simulation
    in the Scarff-II potential (\ref{e:ScarffII}) with $\sigma=-1$, $c=1$ and
    $\epsilon=0.2$. The solid blue line in the left panel is this full
    PDE solution projected onto the phase plane.}
    \label{Fig4}
\end{figure}

When $\sigma_1<0$, the phase portrait is shown in Fig.~\ref{Fig4}
(left panel), where $\alpha$ and $\sigma_1$ values are taken from
Eq. (\ref{e:alphasigma}) with $c=1$ and $\sigma=-1$ (defocusing
nonlinearity). In this case, except for the origin (an unstable
fixed point), all trajectories escape to infinity. Similar solution
behaviors are observed in the full PDE \eqref{Eq:LatticeNLS}. An
example is shown in the right panel of Fig.~\ref{Fig4}, where the
PDE solution is seen to first decrease, and then rise to high
amplitudes. The projection of this PDE solution onto the phase plane
is displayed as a solid blue line in the phase portrait. This
projection closely follows the trajectory of the ODE model. After
the solution amplitude has reached the order $A\sim 1/\epsilon$
(beyond the validity of our perturbation theory), PDE solutions can
eventually saturate in amplitude while continuing to shed radiation
and grow in power.

\section{Solution Behaviors Below Phase Transition}

In this section, we consider the predictions of our reduced model
for solution behaviors below phase transition, and compare them with
PDE solutions.

\subsection{Soliton solutions}

Below phase transition, $\alpha<0$, the ODE model (\ref{Eq:AFull})
admits constant-amplitude solutions (\ref{e:AZ}) for both signs of
the nonlinear coefficient $\sigma_1$, meaning that solitons exist
under both focusing and defocusing nonlinearities. But behaviors of
solitons for the two signs of $\sigma_1$ are very different.

\begin{figure}[!htbp]
    \centering
    \includegraphics[width=0.48\textwidth]{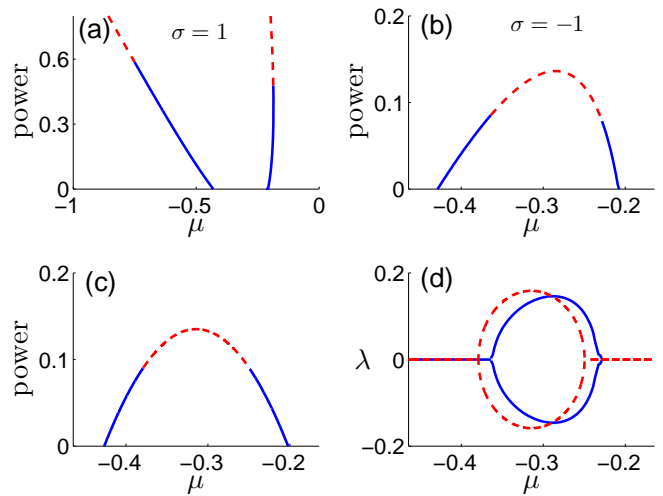}
    \caption{(a,b) Numerically obtained power curves for the families of solitons below phase transition in the Scarff-II potential (\ref{e:ScarffII}) with
     $c=-1$, $\epsilon=0.2$, $\sigma=1$ in (a) and $\sigma=-1$ in (b). (c) Analytical prediction for the power curve and linear stability of solitons in (b).
     In all of (a,b,c), solid blue indicates stable solitons and dashed red unstable ones. (d) Comparison of numerically obtained (solid blue) and
     analytically predicted (dashed red) unstable eigenvalues for solitons in (b). }
    \label{Fig5}
\end{figure}

When $\sigma_1>0$, formula (\ref{e:mu1}), when rewritten as
\begin{equation} \label{e:A02b}
A_0^2 = (\mu_1^2+\alpha)/\sigma_1,
\end{equation}
predicts that constant-amplitude solutions exist when $|\mu_1|>
\sqrt{|\alpha|}$, i.e., soliton solutions exist when
$|\mu-\mu_0|>\epsilon \sqrt{|\alpha|}$. In addition, the amplitude
$A_0$ (and hence power) of these solitons can be arbitrary.
Numerically we have confirmed this prediction in the Scarff-II
potential (\ref{e:ScarffII}) with $\sigma=1$, $c=-1$ and
$\epsilon=0.2$. The numerically obtained power curves of these
solitons are displayed in Fig.~\ref{Fig5}(a). Stability of these
solitons can be analyzed in the framework of the reduced model
(\ref{Eq:AFull}), and the eigenvalue formula (\ref{e:lambda}) shows
that these solitons are linearly stable, which agrees with the
numerical findings in Fig.~\ref{Fig5}(a) for solitons at lower
amplitudes (where the perturbation theory is valid). At higher
amplitudes, the solitons do become unstable, similar to the case
above phase transition in Fig.~\ref{Fig2}(a).

When $\sigma_1<0$, formulae (\ref{e:mu1}) and (\ref{e:A02b}) predict
that solitons only exist in the propagation-constant interval of
$|\mu-\mu_0|<\epsilon \sqrt{|\alpha|}$ with a limited range of
amplitude values $|A_0|\le \sqrt{\alpha/\sigma_1}$. The analytically
predicted power curve from Eqs. (\ref{e:psiexpansion}),
(\ref{e:u1u2}) and (\ref{e:A02b}) is
\begin{equation}
P(\mu)\approx \frac{(\mu-\mu_0)^2+\alpha \epsilon^2}{\sigma_1} \int_{-\infty}^\infty |u_e|^2dx,
\end{equation}
which is plotted in Fig.~\ref{Fig5}(c). Numerically we have obtained
these solitons, whose power curve is shown in Fig.~\ref{Fig5}(b).
This numerical power curve closely resembles the analytical
prediction in (c). In particular, the existence of a power upper
bound is confirmed. This close agreement between the perturbation
theory and direct numerics is understandable, since these solitons
have low powers and are thus within the regime of validity of the
perturbation theory.

The physical reason for limited power ranges of these solitons is
that, under defocusing nonlinearity, if this power is too large, the
negative nonlinearity-induced refractive index would transform the
effective potential from below phase transition to above phase
transition, rendering stationary solitons impossible.

Stability of these solitons with limited power ranges can be
analyzed in the framework of the reduced model (\ref{Eq:AFull}). In
this case, the eigenvalue formula (\ref{e:lambda}) predicts that
these solitons are linearly unstable when
\begin{equation}
|\mu-\mu_0| < \epsilon \sqrt{|\alpha|/3},
\end{equation}
and stable otherwise. This predicted instability and stability is
shown on the predicted power curve in Fig.~\ref{Fig5}(c). It is seen
that solitons at the top part of the power curve are predicted as
unstable and the bottom ones predicted as stable. Numerically we
have determined the linear stability of these solitons by computing
their stability spectra, and the results are shown in
Fig.~\ref{Fig5}(b). Clearly the numerical results match those of
analytical predictions. Quantitatively we have also computed real
eigenvalues of unstable solitons and plotted them in
Fig.~\ref{Fig5}(d), together with their analytical predictions in
Eq. (\ref{e:lambda}). This quantitative comparison shows good
agreement as well.

\subsection{Oscillating solutions}

Like the previous case above phase transition, robust oscillating
solutions exist below phase transition as well. As before, we will
unveil such solutions by focusing on the case of real $A$ in the
reduced model \eqref{Eq:AFull}.

\begin{figure}[!htbp]
    \centering
    \includegraphics[width=0.48\textwidth]{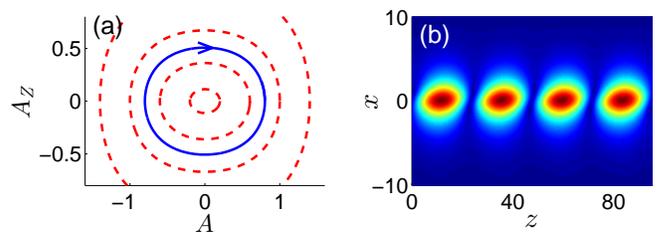}
    \caption{(Left) Phase portrait of the reduced model \eqref{Eq:AFull} with $\sigma_1>0$ below phase transition. (Right) Full PDE simulation
    in the Scarff-II potential (\ref{e:ScarffII}) with $\sigma=1$, $c=-1$ and
    $\epsilon=0.2$. The solid blue line in the left panel is this full
    PDE solution projected onto the phase plane.}
    \label{Fig6}
\end{figure}

If $\sigma_1>0$, the phase portrait of the reduced model is shown in
Fig. \ref{Fig6} (left panel), where $\alpha$ and $\sigma_1$ values
are taken from Eq. (\ref{e:alphasigma}) with $c=-1$ and $\sigma=1$
(focusing nonlinearity). In this phase portrait the origin is a
stable fixed point, a reflection that the system is below phase
transition. Surrounding the origin are periodic orbits of various
sizes. This implies an abundance of robust oscillating solutions in
the PDE system. Numerically we have confirmed the existence of these
oscillating solutions, and an example is shown in Fig. \ref{Fig6}
(right panel). Projection of this PDE solution onto the phase plane
is plotted as a solid blue line in the left panel, and good
agreement with the ODE orbit is seen.

\begin{figure}[!htbp]
    \centering
    \includegraphics[width=0.48\textwidth]{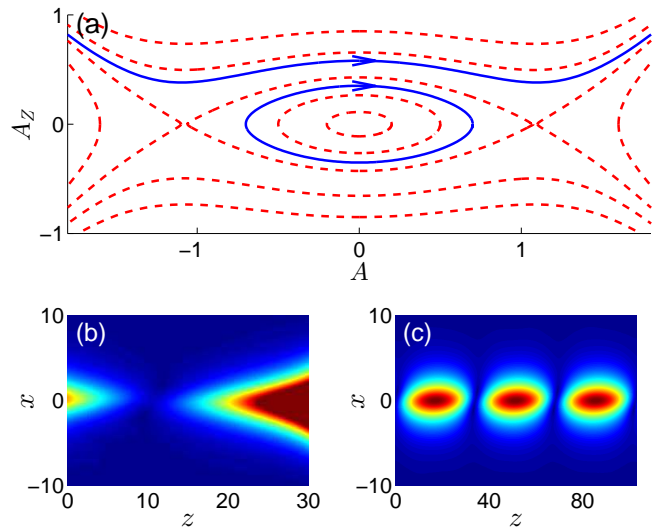}
    \caption{(a) Phase portrait of the reduced model \eqref{Eq:AFull} with $\sigma_1<0$ below phase transition. (b,c) Full PDE simulations
    in the Scarff-II potential (\ref{e:ScarffII}) with $\sigma=-1$, $c=-1$ and $\epsilon=0.2$ under different initial conditions. In (a), solid blue lines are full
    PDE solutions in (b,c) projected onto the phase plane [with the upper curve for (b) and lower curve for (c)].}
    \label{Fig7}
\end{figure}

If $\sigma_1<0$, the phase portrait of the reduced model
\eqref{Eq:AFull} is shown in Fig. \ref{Fig7}(a), where $\alpha$ and
$\sigma_1$ values are taken from Eq. (\ref{e:alphasigma}) with
$c=-1$ and $\sigma=-1$ (defocusing nonlinearity). This phase
portrait contains three fixed points: the origin which is stable,
and $A=\pm \sqrt{\alpha/\sigma_1}$ which are unstable. The latter
two fixed points correspond to the soliton with maximal power (at
$\mu=\mu_0$) in Fig.~\ref{Fig5}(c). Away from these three
equilibria, trajectories are divided into two categories: periodic
orbits surrounding the origin, and orbits which escape to infinity.
Numerically we have found both types of solutions in the PDE system
\eqref{Eq:LatticeNLS} under the Scarff-II potential
(\ref{e:ScarffII}) with $\sigma=-1$, $c=-1$ and $\epsilon=0.2$, and
two examples are displayed in Fig. \ref{Fig7}(b,c). Projections of
the PDE solutions onto the phase plane in panel (a) indicate that
the ODE model accurately describes the PDE dynamics.

\section{Summary and Discussion}
In this article, nonlinear wave propagation in $\PT$-symmetric
localized potentials was investigated analytically near phase
transition. Necessary conditions for phase transition were first
derived based on a generalization of the Krein signature. Then rich
nonlinear dynamics near phase transition was revealed through a
multi-scale perturbation analysis, which yielded a nonlinear ODE
model for the amplitude of the solutions. Above phase transition,
this ODE model predicted a family of stable solitons not bifurcating
from linear modes under a certain sign of nonlinearity. In addition,
it predicted persistent periodically-oscillating solutions away from
solitons. Under the opposite sign of nonlinearity, it predicted
unbounded growth of solutions. Below phase transition, solution
dynamics was predicted as well. We have compared all analytical
predictions with direct numerical calculations of the full PDE
system and good agreement was obtained.

The analytical results obtained in this article are helpful for
several reasons. First, it is known that phase transition is a
distinct and important phenomenon in $\PT$-symmetric systems. Thus
the analytical condition for phase transition in terms of
$\PT$-Krein signatures helps understand when phase transition can or
cannot occur. Second, the analytical predictions of nonlinear
dynamics near phase transition contribute to a global understanding
of solution behaviors in $\PT$-symmetric systems. Thirdly, even
though our analysis was performed only for the potential NLS
equation (\ref{Eq:LatticeNLS}), a similar treatment can obviously be
extended to other $\PT$-symmetric systems near phase transition, and
similar solution dynamics is expected in all such systems.

\section*{Acknowledgment} This work was supported in part by the Air Force Office of
Scientific Research (USAF 9550-12-1-0244) and the National Science
Foundation (DMS-1311730).

\bigskip

\end{document}